%% file: main.tex
\documentclass[12pt,a4paper]{article}
\pdfoutput=1
\usepackage{ifthen} % for conditional statements
\usepackage{ragged2e}
\usepackage{multirow}
\newboolean{pdflatex}
\setboolean{pdflatex}{true} % False for eps figures 

\newboolean{articletitles}
\setboolean{articletitles}{true} % False removes titles in references

\newboolean{uprightparticles}
\setboolean{uprightparticles}{false} %True for upright particle symbols

\newboolean{inbibliography}
\setboolean{inbibliography}{false} %True once you enter the bibliography

\input{preamble}

\usepackage{longtable} % only for template; not usually to be used in PAPERs
\usepackage[colorinlistoftodos,prependcaption,textsize=tiny]{todonotes}

\begin{document}

%%%%%%%%%%%%%%%%%%%%%%%%%
%%%%% Title     %%%%%%%%%
%%%%%%%%%%%%%%%%%%%%%%%%%
\renewcommand{\thefootnote}{\fnsymbol{footnote}}
\setcounter{footnote}{1}
%\input{our-symbols-def}

%\onecolumn
\input{titlepage}

\newcommand{\tmpdg}[1]{{\textcolor{red}{[DG: #1]}}}
\renewcommand{\thefootnote}{\arabic{footnote}}
\setcounter{footnote}{0}

\pagestyle{plain} % restore page numbers for the main text
\setcounter{page}{1}
\pagenumbering{arabic}

%% Uncomment during review phase. 
%% Comment before a final submission.
%\linenumbers
% \listoftodos
\input{body}

\addcontentsline{toc}{section}{References}
\setboolean{inbibliography}{true}
\bibliographystyle{fd}
\bibliography{main,LHCb-PAPER}

\end{document}

%% file: preamble.tex
\usepackage[top=1in, bottom=1.25in, left=1in, right=1in]{geometry}

\usepackage{microtype}
\usepackage{lineno}  % for line numbering during review
\usepackage{xspace} % To avoid problems with missing or double spaces after
\usepackage{caption} %these three command get the figure and table captions automatically small

\usepackage{graphicx}  % to include figures (can also use other packages)
\usepackage{color}
\usepackage{colortbl}
\graphicspath{{./figs/}} % Make Latex search fig subdir for figures

\usepackage{amsmath} % Adds a large collection of math symbols
\usepackage{amssymb}
\usepackage{amsfonts}
\usepackage{booktabs}
\usepackage{upgreek} % Adds in support for greek letters in roman typeset

\newcommand*\patchAmsMathEnvironmentForLineno[1]{%
\expandafter\let\csname old#1\expandafter\endcsname\csname #1\endcsname
\expandafter\let\csname oldend#1\expandafter\endcsname\csname
end#1\endcsname
 \renewenvironment{#1}%
   {\linenomath\csname old#1\endcsname}%
   {\csname oldend#1\endcsname\endlinenomath}%
}
\newcommand*\patchBothAmsMathEnvironmentsForLineno[1]{%
  \patchAmsMathEnvironmentForLineno{#1}%
  \patchAmsMathEnvironmentForLineno{#1*}%
}
\AtBeginDocument{%
\patchBothAmsMathEnvironmentsForLineno{equation}%
\patchBothAmsMathEnvironmentsForLineno{align}%
\patchBothAmsMathEnvironmentsForLineno{flalign}%
\patchBothAmsMathEnvironmentsForLineno{alignat}%
\patchBothAmsMathEnvironmentsForLineno{gather}%
\patchBothAmsMathEnvironmentsForLineno{multline}%
\patchBothAmsMathEnvironmentsForLineno{eqnarray}%
}

\usepackage{hyperref}    % Hyperlinks in references
\usepackage[all]{hypcap} % Internal hyperlinks to floats.

\input{our-symbols-def}

\input{our-defs}

\usepackage{cite} % Allows for ranges in citations
\usepackage{mciteplus}

\usepackage{xcolor} % [dvipsnames] 
\definecolor{halfgray}{gray}{0.55} % chapter numbers will be semi transparent .5 .55 .6 .0
\definecolor{webgreen}{rgb}{0,.5,0}
\definecolor{webbrown}{rgb}{.6,0,0}
\definecolor{Maroon}{cmyk}{0, 0.87, 0.68, 0.32}
\definecolor{RoyalBlue}{cmyk}{1, 0.50, 0, 0}

\hypersetup{%
    colorlinks=true, linktocpage=true, pdfstartpage=3, pdfstartview=FitV,%
    breaklinks=true, pdfpagemode=UseNone, pageanchor=true, pdfpagemode=UseOutlines,%
    plainpages=false, bookmarksnumbered, bookmarksopen=true, bookmarksopenlevel=1,%
    urlcolor=webbrown, linkcolor=RoyalBlue, citecolor=webgreen, %pagecolor=RoyalBlue,%
    pdftitle={Tracking the B},%
    pdfauthor={},%
    pdfsubject={},%
    pdfkeywords={},%
    pdfcreator={pdfLaTeX},%
  }

\usepackage{wrapfig}

%% file: our-symbols-def.tex
\definecolor{heartgold}{rgb}{0.5, 0.5, 0.0}

 \def\PJ      {\ensuremath{J}\xspace}

 \def\Pi      {\ensuremath{i}\xspace}

\def\Ppsi {\ensuremath{\psi}\xspace}
\def\jpsi     {{\ensuremath{{\PJ\mskip -3mu/\mskip -2mu\Ppsi\mskip 2mu}}}\xspace}
\def\Jpsi     {{\ensuremath{\jpsi}}\xspace}

\def\Bbar    {\kern 0.18em\overline{\kern -0.18em B}{}\xspace}

\newcommand{\Bu}{\ensuremath{B^+}\xspace}
\newcommand{\Bc}{\ensuremath{B^+_c}\xspace}

\newcommand{\B}{\ensuremath{B}\xspace}

\mathchardef\PLambda="7103

\def\Kbar  {\kern 0.2em\overline{\kern -0.2em K}{}\xspace}

\newcommand{\tev}{\ifthenelse{\boolean{inbibliography}}{\ensuremath{~T\kern -0.05em eV}\xspace}{\ensuremath{\mathrm{\,Te\kern -0.1em V}}}\xspace}
\newcommand{\gev}{\ensuremath{\mathrm{\,Ge\kern -0.1em V}}\xspace}
\newcommand{\mev}{\ensuremath{\mathrm{\,Me\kern -0.1em V}}\xspace}
\newcommand{\kev}{\ensuremath{\mathrm{\,ke\kern -0.1em V}}\xspace}
\newcommand{\ev}{\ensuremath{\mathrm{\,e\kern -0.1em V}}\xspace}
\newcommand{\gevc}{\ensuremath{{\mathrm{\,Ge\kern -0.1em V\!/}c}}\xspace}
\newcommand{\mevc}{\ensuremath{{\mathrm{\,Me\kern -0.1em V\!/}c}}\xspace}
\newcommand{\gevcc}{\ensuremath{{\mathrm{\,Ge\kern -0.1em V\!/}c^2}}\xspace}
\newcommand{\gevgevcccc}{\ensuremath{{\mathrm{\,Ge\kern -0.1em V^2\!/}c^4}}\xspace}
\newcommand{\mevcc}{\ensuremath{{\mathrm{\,Me\kern -0.1em V\!/}c^2}}\xspace}

\newcommand{\pip}{\ensuremath{\pi^+}\xspace}

\def\cm   {\ensuremath{\mathrm{ \,cm}}\xspace}

\def\mm   {\ensuremath{\mathrm{ \,mm}}\xspace}

%% file: our-defs.tex
\def\pizero{\ensuremath{\pi^0}\xspace}

\def\bujpsik{\ensuremath{B^+ \to J/\psi K^+}\xspace}

\def\ks{\ensuremath{K^0_S}\xspace}

\def\ks{\ifmmode \mr{K^0_S} \else $\mr{K^0_S}$\xspace\fi}

\def\bujpsik{\ensuremath{B^+ \to J/\psi K^+}\xspace}

\def\buellnu{\ensuremath{B^+ \to \ell^+ \nu}\xspace}
\def\Bu{\ensuremath{B^+}\xspace}

%% file: titlepage.tex
%%%%%%%%%%%%%%%%%%%%%%%%%
%%%%%  TITLE PAGE  %%%%%%
%%%%%%%%%%%%%%%%%%%%%%%%%
\begin{titlepage}

% Header ---------------------------------------------------
\vspace*{-1.5cm}

\noindent

\vspace*{4.0cm}

% Title --------------------------------------------------
{\normalfont\bfseries\boldmath\huge
\begin{center}
  Tracking charged $b$-hadrons:\\
  feasibility study of the use \\of inner trackers \\ to improve $B^+_{(c)}$ reconstruction
\end{center}
}

\vspace*{2.0cm}

% Authors -------------------------------------------------
\begin{center}
F. Dettori$^{1,2}$, A. Lampis$^2$, M. Mulder$^{3,4}$, G. Onderwater$^{4,5}$, D. Provenzano$^{1,2}$, M. van Veghel$^{4,5}$
  \bigskip\\
{\normalfont\itshape\footnotesize
$^1$Universit\`{a} degli Studi di Cagliari  Cagliari, Italy\\
$^2$INFN Cagliari, Cagliari, Italy\\
$^3$Van Swinderen Institute, University of Groningen, Groningen, Netherlands\\
$^4$Nikhef National Institute for Subatomic Physics, Amsterdam, Netherlands\\
$^5$Universiteit Maastricht, Maastricht, Netherlands\\
}
\end{center}

\vspace{\fill}

% Abstract -----------------------------------------------
\begin{abstract}
  \noindent
A method to improve the reconstruction of charged $b$-hadron decays is proposed that uses energy deposits left by the hadron in tracking detectors close to the production point.  Performances are shown for different detector configurations and different number of deposits reconstructed, as obtained in  simulation, for $b$-hadrons produced in high energy proton-proton collisions. 
It is shown that up to few percent of the   
\Bu mesons could leave two deposits before decaying, depending on the detector configuration. 
The presented results can inform the design of future inner detectors. 
This method could increase significantly the physics reach of flavour physics at hadron colliders, opening it to decays with missing particles and vertex information that are otherwise unreconstructable.
\end{abstract}

\vspace*{2.0cm}
\vspace{\fill}

\end{titlepage}

\pagestyle{empty}  % no page number for the title 

%%%%%%%%%%%%%%%%%%%%%%%%%%%%%%%%
%%%%%  EOD OF TITLE PAGE  %%%%%%
%%%%%%%%%%%%%%%%%%%%%%%%%%%%%%%%

%  empty page follows the title page ----
\newpage
\setcounter{page}{2}
\mbox{~}

\cleardoublepage

%% file: body.tex
\section{Introduction}

The physics of heavy flavours, particularly of $b$-hadrons, is a fantastic probe for our understanding 
of fundamental interactions, and particularly of the violation of Charge-Parity (CP) symmetry or searches for new phenomena in rare decays. 
The study of partially reconstructed $b$-hadron final states poses challenges that limit their sensitivity compared to fully reconstructed ones.
This is particularly true at hadron colliders where there is no possibility of closing the kinematics with information from the initial state, 
or where reconstructing the second accompanying $b$-hadron would severely limit the statistics. 
In particular for charged $B$ mesons, i.e. \Bu or \Bc, some final states present minimal information reconstructible at experiments such as only one charged track and one 
or no calorimetric deposit from neutral particles. Among these are rare decays such as \buellnu where $\ell$ is a lepton~\cite{Lees:2012ju,Kronenbitter:2015kls}, sensitive to extensions of the Standard Model with charged heavy particles~\cite{Baek:1999ch}, or CP-violation sensitive decays such as $\Bu \to \pip \pizero$~\cite{BaBar:2007uoe,Belle-II:2023ksq}.
Increasing the background suppression of these and other partially reconstructed decays can enhance the performances
of several experiments as well as opening new ways of looking for new physics. 

In this article  a new method is proposed to search for partially reconstructed decays of charged $b$-hadrons, 
by tracking the hadron before its decay in detectors placed very close to the hadron production vertex. 
Given the typical lifetime of $b$-hadrons~\cite{PDG}, 
when produced in high energy collisions, e.g. at the LHC, their boost allows them to fly up to few centimetres before decaying. 
A few of these hadrons could reach one or more tracking stations, depending on the detector geometry. 
Subsequently, the reconstructed track segment can be used to constrain the direction, and thus the momentum of the $b$ hadron, 
constraining its kinematics in case of missing final state particles, particularly in the case in which no other vertex reconstruction is possible. 
This in turn can pave the way for a new set of measurements or improve their precision, such as semileptonic and other partially reconstructed decays. 
The proposed method can make measurements viable typically thought to be infeasible at hadron colliders that are based on decay channels with limited possibilities of reconstruction of the final state. 

This article stems from the experience of the authors with the LHCb experiment and as such takes this experiment, 
and in particular the Vertex Locator (VELO) as an example; however, the method is more general and not limited to this experiment. 
The use of tracks of long-lived particles in the LHCb experiment, in the case of charged strange hadrons, was discussed in Refs.~\cite{Contu:1693666,LHCb:2024wvl}. In that case, the hadron can fly up to 1 meter and traverse a significant portion of the VELO detector, traversing multiple sections. Standard tracking was exploited and the strange hadron track was connected to its origin vertex. 
Here it is proposed to exploit starting from a single energy deposit (hit) in the tracking detectors to improve the background rejection and kinematic reconstruction of charged \B decays. 
Adding hits can improve the identification of the relevant primary vertex among many, the direction of the \Bu momentum, and the background rejection from neutral \B decays. We therefore consider one to three or more hits, and evaluate the efficiency to reach those stations for different realistic geometry configurations.    
The results here presented can inform the design of future inner tracking devices. 

A full performance study for specific cases would require to take into account a simulation of the whole background, which is the subject of future work. In this article only the signal reconstruction feasibility is considered.

The proposed method relies on tracking detectors very close to the interaction point where the \B hadrons are produced. 
Such subdetectors are present in current particle physics experiments at LHC~\cite{Collaboration:1624070},~\cite{Collaboration:2745805} and \cite{CERN-LHCC-2017-005},
to identify and distinguish primary vertices, tracks and displaced particles.
In this paper we take as benchmarks the VELO~\cite{LHCbVELOGroup:2014uea} detector of the LHCb experiment and its current and future upgrades. However, results can be applied to any detector with similar distance to the interaction point with minor changes.
In the following we will consider a generic tracking detector with these configurations, measuring 2D coordinates of particles at fixed $z$ distances regardless of the hardware technology that allows this. 

The paper is organised as follows: in \S~\ref{sec:geometry} we present the considered detector geometries, the \Bu simulation and tracking feasibility is presented in \S~\ref{sec:simulation}, while in \S~\ref{sec:results} we discuss the results and draw our conclusions.

\section{Geometry of the detectors }
\label{sec:geometry}

\begin{figure}[tbp]
\begin{center}
 \includegraphics[width = 0.45\textwidth]{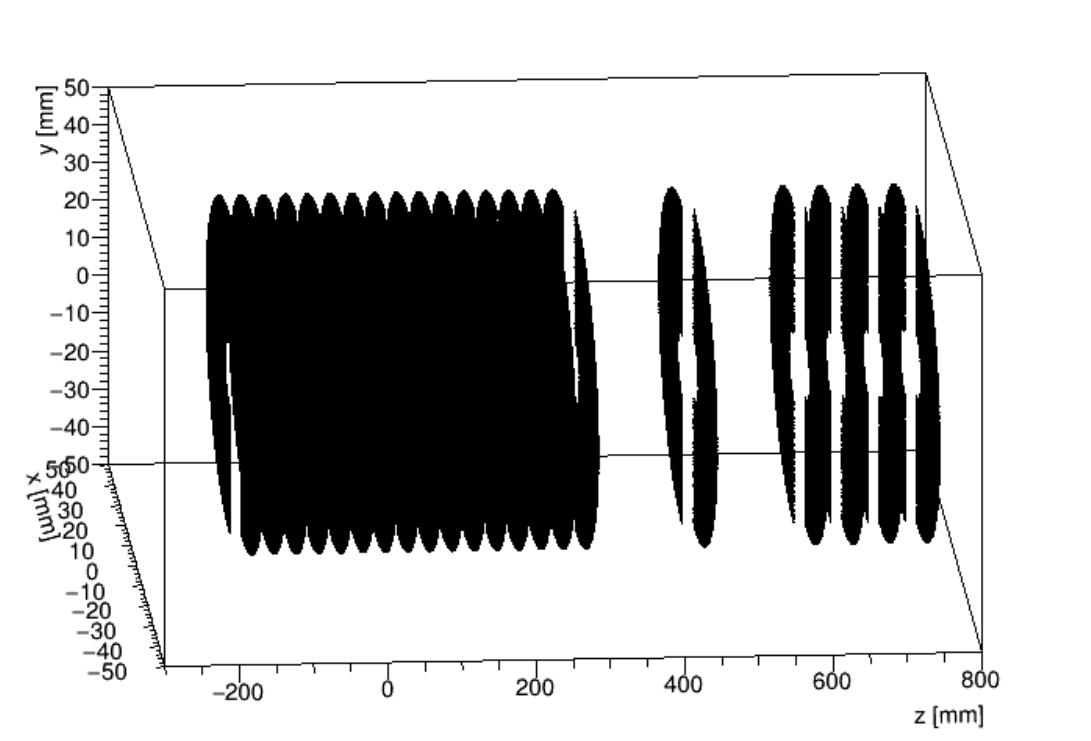}
  \includegraphics[width = 0.45\textwidth]{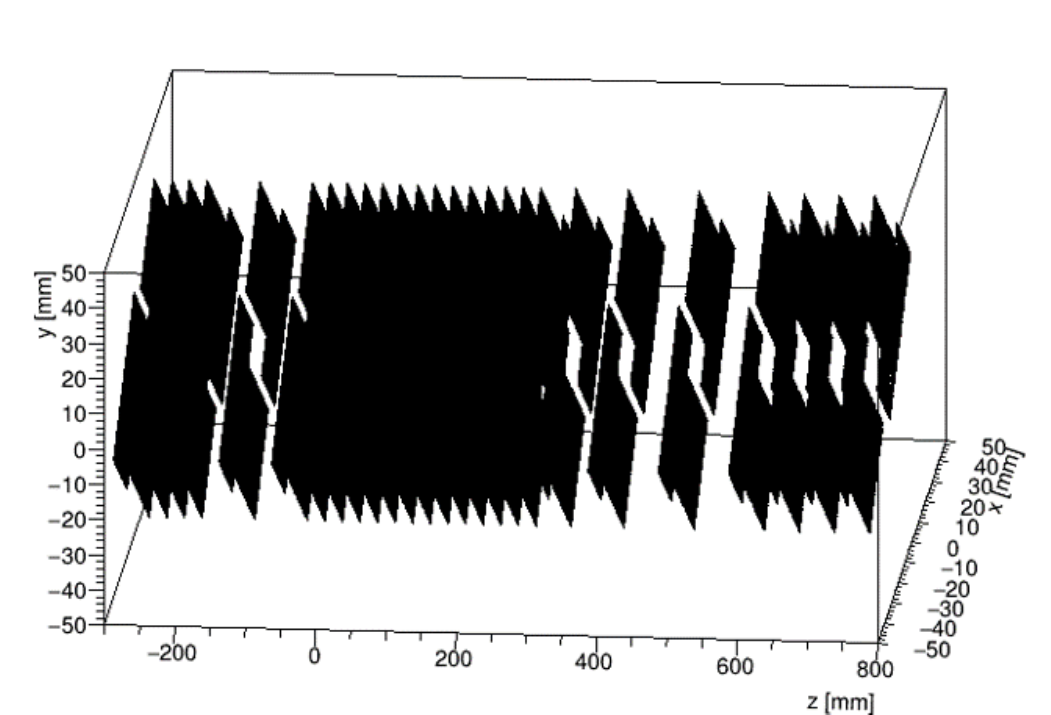}
  \caption{\emph{Examples of the simulated geometries.(Left) Run~1/2 VELO; (Right) Run~3/4 VELO.}}\label{fig:geo}
  \end{center}
 \end{figure}

Six different geometries are considered in this paper, all of them built with custom software. Two example geometries are shown in Fig.~\ref{fig:geo}.
Each detector configuration is composed of several identical tracking stations. For the studied geometries, four types of tracking stations are considered. To simplify the geometries, the stations are simulated with an infinitesimal thickness, since the sensor thickness is negligible with respect to the typical \Bu flight distances considered, and multiple scattering is in any case not simulated. 

The detector configurations used as case studies are chosen based on different designs of the LHCb VELO detector, both past and future.
The VELO is located around the proton-proton interaction point and provides the first tracking points for charged particles, 
allowing precise reconstruction of primary and secondary vertices. 
The VELO is a silicon detector composed of a series of tracking sensors in the transverse plane positioned along the beam direction. 
This study covers the former VELO used in Run 1 and 2 between 2010 and 2018 which uses silicon microstrips, the new VELO detector based on silicon pixel technology, installed for Run 3 and Run 4, and new scenarios which are being considered in view of future LHCb upgrades.

Because of the aforementioned technologies, during LHC Run 1 and 2, the VELO detector measured $r$ and $\phi$ coordinates (in a cylindrical coordinate system where the $z$ axis is the beam direction), while the current upgraded detector measures Cartesian coordinates in the transverse plane. 
The distance between two consecutive stations is 3~\cm (2.5~\cm) in Run~1/2 (Run~3/4); as shown in the following, this is a major parameter in the efficiency of this method. 
The distance of the sensors from the beam-line is 8~\mm (5.1~\mm)  in Run~1/2 (Run~3/4).
Thus, the typical radial distance of the first hit 
of a charged particle decreases from 10~\mm to about 6~\mm. 
The Run 1/2 geometry is composed of 42 tracking stations. The stations are perpendicular to the beam direction and spaced out at 30 mm intervals. Each tracking station is a semi-circle with a radius of 42 mm. The inner 8.2 mm radius semi-circle is not sensitive. 
The Run~3/4 geometry has 52 equal tracking stations transverse to the beam direction and spaced 25 mm from each other. Each tracking station has an L shaped sensitive area consisting of two adjacent rectangles of $42.57\mm \times 28\mm$. The detector sensitive areas are placed at 5.1 mm from the beam line at the closest point.

These first two geometries are a representation of two previously or currently existing detectors; four more geometries were developed to test the method in different conditions. Two of them try to maximise performance during possible LHC Run 5 and 6 scenarios, and two are similar to the proposed detector in the LHCb Upgrade II FTDR~\cite{CERN-LHCC-2021-012}.
The \emph{Closer} geometry is equal to the Run~3/4 one but with a reduced distance from the beam line from 5.1 \mm to 4 \mm. This distance is chosen such that the radiation fluence to the innermost sensors is equal to $1\cdot10^{17}\rm{n_{eq}}\cm^{-2}$, a value for which 3D trench sensors, proposed as a candidate technology for the Phase II Upgrade of the LHCb VELO detector, have demonstrated full efficiency~\cite{10.3389/fphy.2024.1497267}.
The \emph{B scenario} and the \emph{X scenario} are geometries presented in the LHCB FTDR for the Phase II Upgrade that are similar to the Run~3/4 VELO geometry but with an increase distance from the beam line up to 12.5 mm (7.1 mm) for the B scenario (X Scenario).
The last simulated geometry is named \emph{Double}, and consists of a Run~3/4 geometry but with double the number of stations, so that the distance between the active planes is reduced from 25 mm to 12.5 mm.

A summary of the considered geometries with their characteristics is shown in Table~\ref{tab:geometry}.

It has to be noted that a proper evaluation of such variations in the design of a real detector will have to take into account much more than what can be  discussed here, in particular multiple scattering and in general radiation lengths, as well as data rate. 

\begin{table}[tb!]
\centering
    \caption{
    Geometry configuration of the considered detectors}
    \label{tab:geometry}
    \vspace{0.3cm}
\begin{tabular}{l| c cc  }
\toprule
Name & Geometry & Distance between stations & Distance to beam line \\
\midrule
Run 1/2  & Semi-circular & 30.0 mm & 8.2 mm \\%
Run~3/4 & L shape & 25.0 mm & 5.1 mm\\
Run 5/6 Closer & L shape & 25.0 mm & 4.0 mm  \\%
Run 5/6 Double & L shape & 12.5 mm & 5.1 mm \\ 
Run 5/6 B Scenario & L shape & 25.0 mm & 12.5 mm\\
Run 5/6 X Scenario & L shape & 25.0 mm & 7.1 mm\\
\bottomrule
\end{tabular}
\end{table}

\section{Simulation of \Bu decays and tracking feasibility}
\label{sec:simulation}

For this study, simulated \Bu decays produced in proton-proton ($pp$) collisions are employed. 
The generation of $pp$ collisions is performed with Pythia~\cite{Sjostrand:2007gs}, 
tuned to the configuration in Ref.~\cite{LHCb-PROC-2010-056}, 
and decays of \B mesons are generated with EvtGen~\cite{Lange:2001uf}.  

A sample of 4 million $pp$ collisions at 13~\tev is generated,
in which \Bu mesons are produced and forced to decay to the \bujpsik final state; note that the particular final state is mostly irrelevant for the performance of the \Bu tracking, as that happens before decay.

\begin{figure}
\begin{center}
\includegraphics[width = 0.9\textwidth]{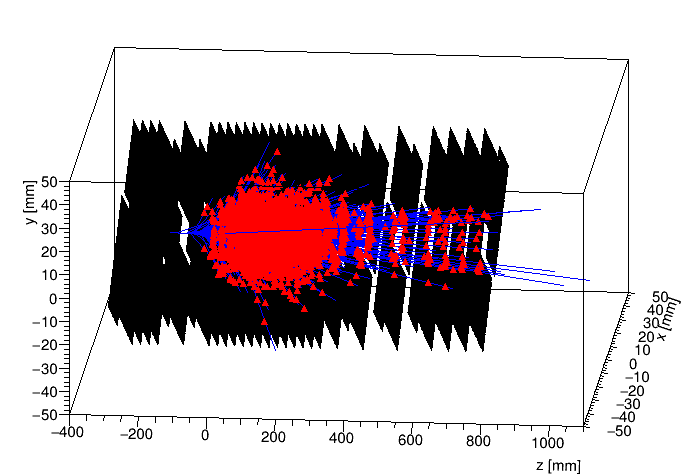}
\includegraphics[width = 0.9\textwidth]{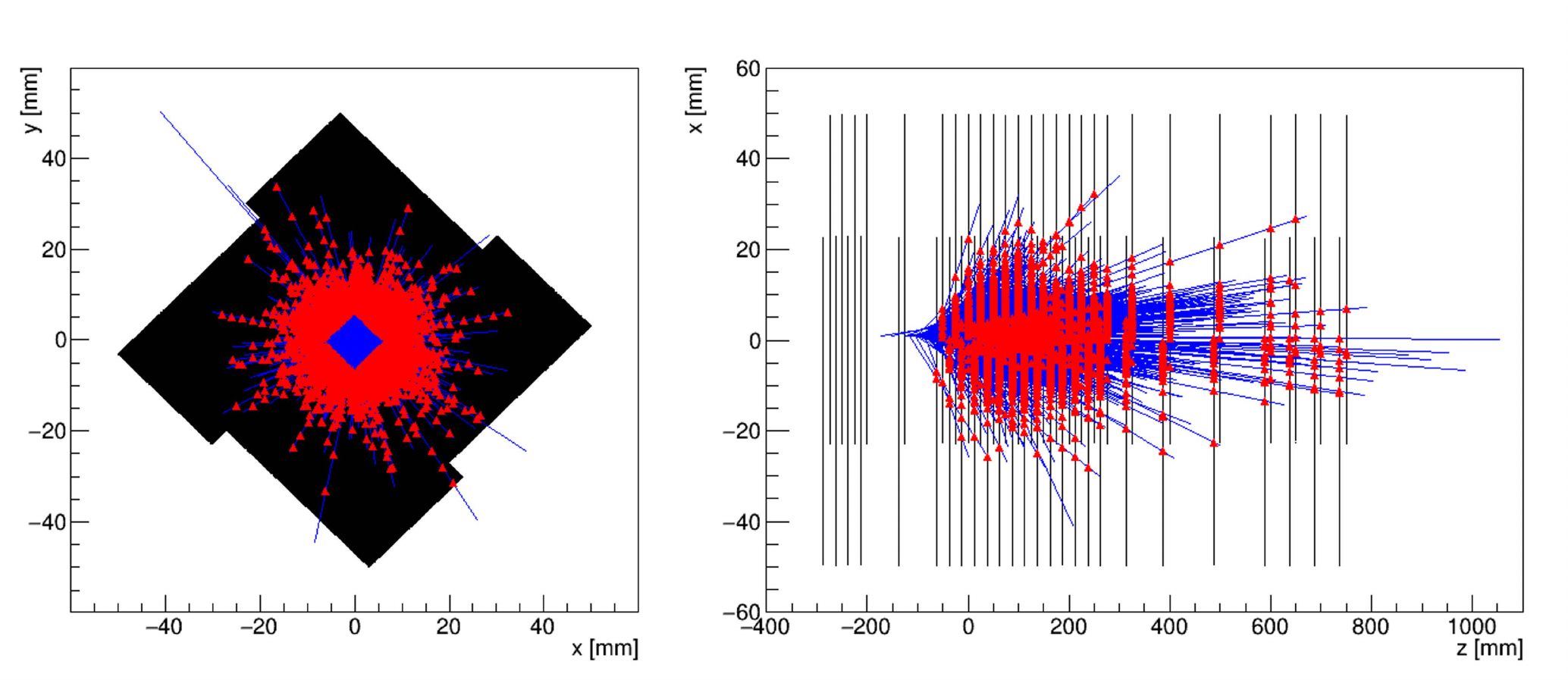}
  \caption{Sample of simulated \Bu mesons in proton-proton collisions at 13 TeV traversing at least two detecting stations in the Run~3/4 geometry. Blue lines represent the paths of the mesons before decaying,   red dots represent energy deposits.   }\label{fig:events}
  \end{center}
 \end{figure}

Only events where the $K^+$ meson and the two muons produced in the decay of the \Jpsi meson are inside the LHCb acceptance are considered in this work. Emulating the LHCb acceptance, these particles are required to be in a pseudorapidity range of $2<\eta<5$. 
The kinematic variables of the generated particles are reported in Appendix~\ref{app} Fig.~\ref{fig:kin} for reference.

Considering each of the detector geometries, the stations crossed by the \Bu meson between its production and decay vertices are counted. 
The detection efficiency of a station in its active area is assumed to be 100\%, which is very similar to the efficiency in real operating conditions~\cite{Aaij:2014zzy}.
A sub-sample of generated events obtained with the Run~3/4   geometry is shown in Fig.~\ref{fig:events}, showing that a significant portion of the \Bu mesons crosses at least two stations before decaying. 
From the number of deposits left by the \Bu mesons in the tracking stations, it is possible to estimate the fraction of \Bu that can be tracked and reconstructed according to thresholds defined below.

 A \Bu meson is considered tracked if it intercepts at least 2 tracking stations, even if a stand-alone track is typically obtained by at least three hits. Even a single hit, together with the position of the primary vertex and in certain cases on subsequent decay vertices, can be used to enhance the reconstruction of decays with missing particles.

\section{Results and discussion}
\label{sec:results}

The fractions of the \Bu mesons that intercept one or more detector stations are shown in Table~\ref{tab:Fraction}.
More than 1 per mille of the mesons have at least 1 hit in the Run~3/4 configuration, and $2.4\cdot 10^{-4}$ have 3 or more deposits.

\begin{table}[b]
\centering
    \caption{
    Fraction of \Bu that produce one or more hits for the studied VELO geometries. Uncertainties are statistical from the sample size.}
    \label{tab:Fraction}
    \vspace{0.3cm}
\begin{tabular}{l| c cc  }
\toprule
\multirow{2}{*}{Geometry} & \multicolumn{3}{c}{\Bu fraction} \\
 & $\geq 1$ hit  & $\geq 2$ hits   & $\geq 3$ hits \\
\midrule
Run 1/2  &  $\left( 4.5 \pm 0.1 \right)\cdot 10^{-4}$ & $\left( 1.28 \pm 0.06 \right)\cdot 10^{-4}$ & $\left( 5.0 \pm 0.4 \right)\cdot 10^{-5}$\\%
Run~3/4 & $\left( 1.80 \pm 0.02 \right)\cdot 10^{-3}$ & $\left( 5.3 \pm 0.1 \right)\cdot 10^{-4}$ & $\left(2.36 \pm 0.08 \right)\cdot 10^{-4}$\\%
Run 5/6 Closer & $\left( 3.77  \pm 0.03 \right)\cdot 10^{-3}$ & $\left( 1.11 \pm 0.02 \right)\cdot 10^{-3}$ & $\left(4.8 \pm 0.1 \right)\cdot 10^{-4}$\\%
Run 5/6 Double & $\left( 2.41  \pm 0.02 \right)\cdot 10^{-3}$ & $\left( 1.15 \pm 0.02 \right)\cdot 10^{-3}$ & $\left(6.4 \pm 0.2 \right)\cdot 10^{-4}$\\%
Run 5/6 B Scenario & $\left( 7.1  \pm 0.4 \right)\cdot 10^{-5}$ & $\left( 2.6 \pm 0.3 \right)\cdot 10^{-5}$ & $\left(1.1 \pm 0.2 \right)\cdot 10^{-5}$\\%
Run 5/6 X Scenario & $\left( 6.0  \pm 0.1 \right)\cdot 10^{-4}$ & $\left( 1.8 \pm 0.07 \right)\cdot 10^{-4}$ & $\left(8.3 \pm 0.4 \right)\cdot 10^{-5}$\\%
\bottomrule
\end{tabular}
\end{table}

\begin{figure}
\begin{center}
 \includegraphics[width = 0.7\textwidth]{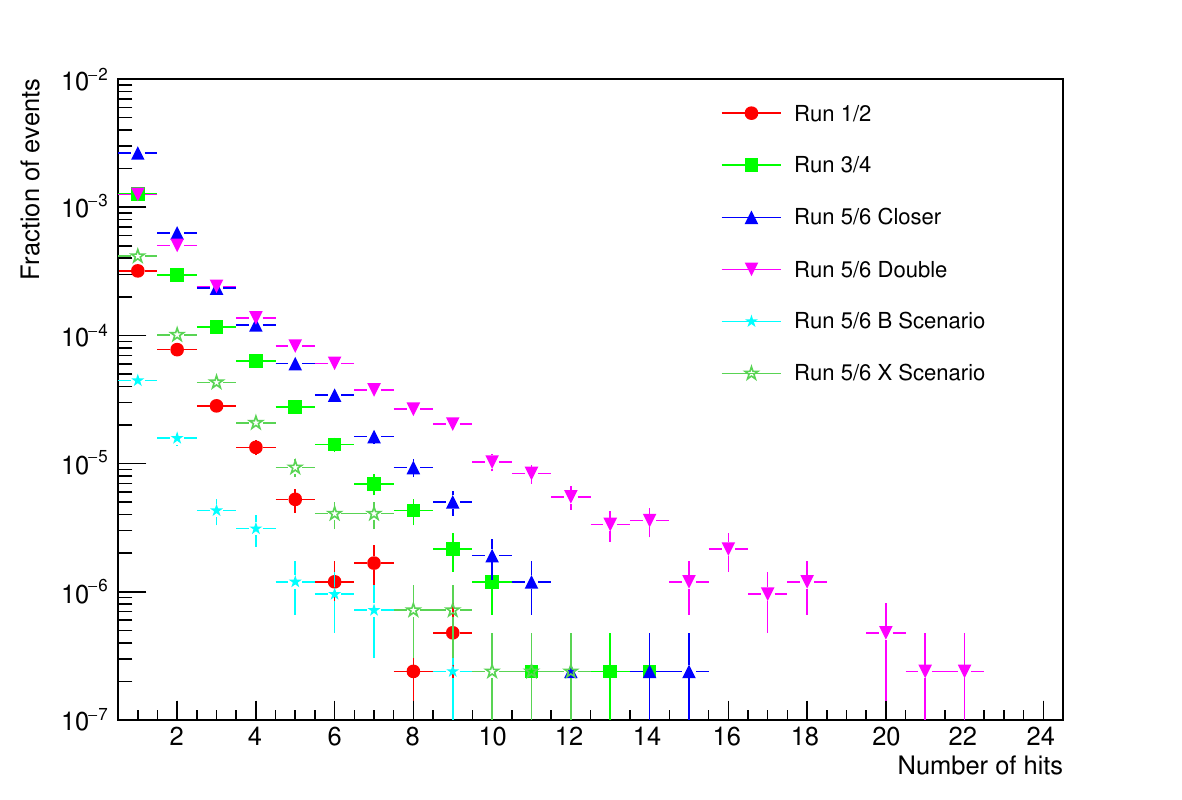}

  \caption{Fraction of \Bu mesons as a function of the number of hits produced for the different geometry configurations.}\label{fig:fraction}
  \end{center}
 \end{figure}

Other geometry configurations show that moving closer to the beam line and decreasing the distance between planes increases the efficiency as expected by meson decay length. The two Run~5/6 Closer and Double geometries allow to increase the fraction of \Bu that can be tracked by a factor two. Reducing the spacing between tracking stations significantly increases  the number of events in which \Bu mesons produce more than two hits, as shown in Fig.~\ref{fig:fraction}.
The B scenario, having a larger distance, appears to be instead disadvantaged. 

\begin{figure}
\begin{center}
 \includegraphics[width = 0.49\textwidth]{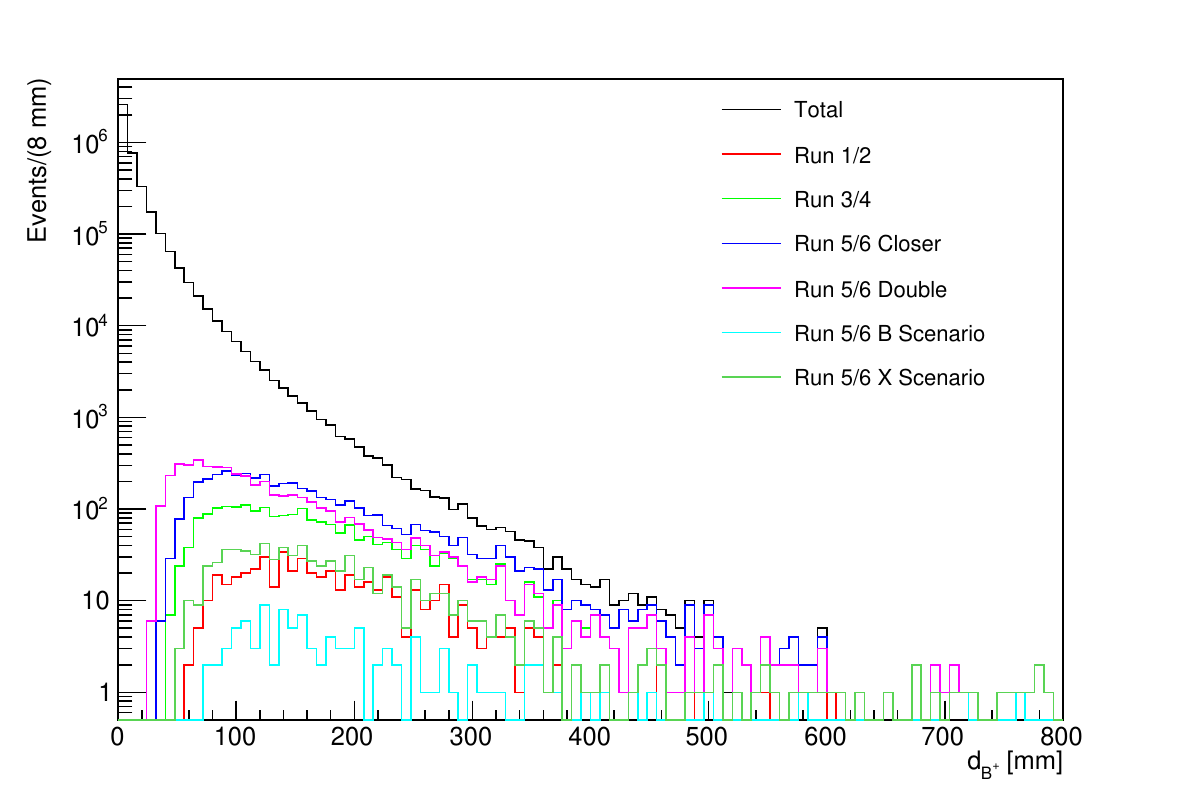}
 \includegraphics[width = 0.49\textwidth]{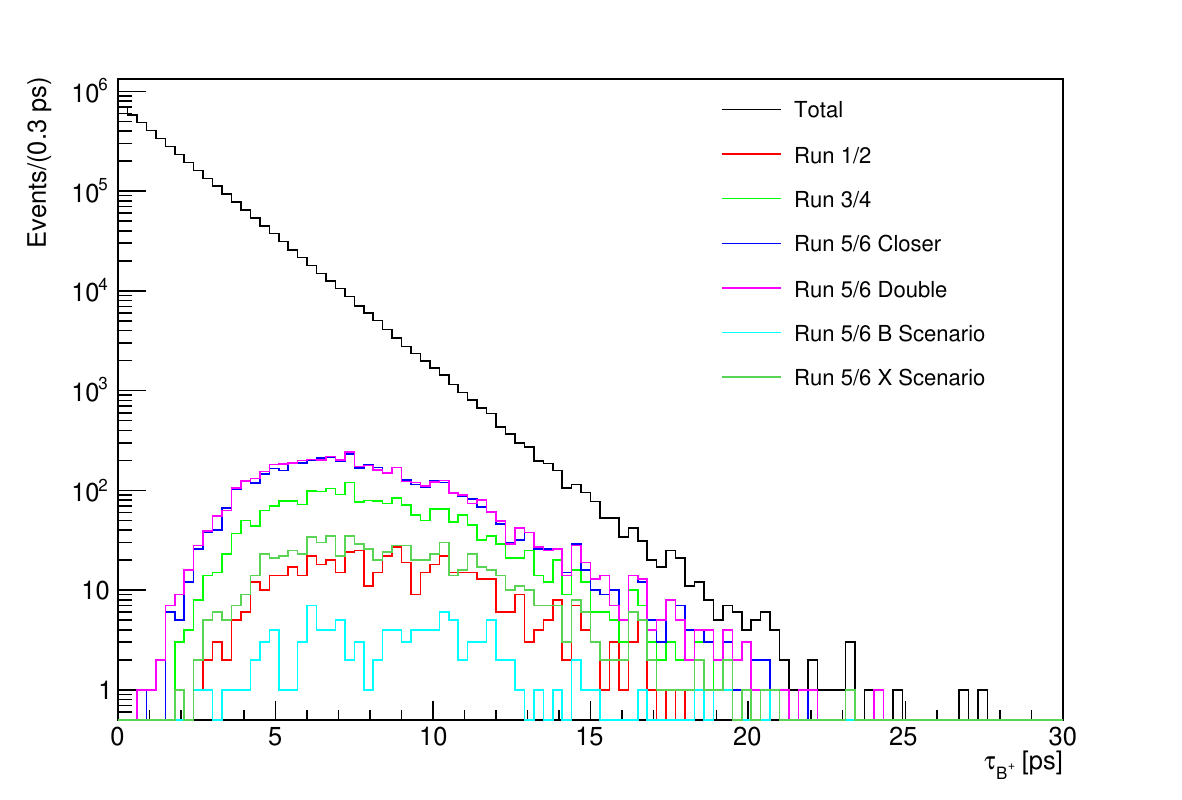}
 \includegraphics[width = 0.49\textwidth]{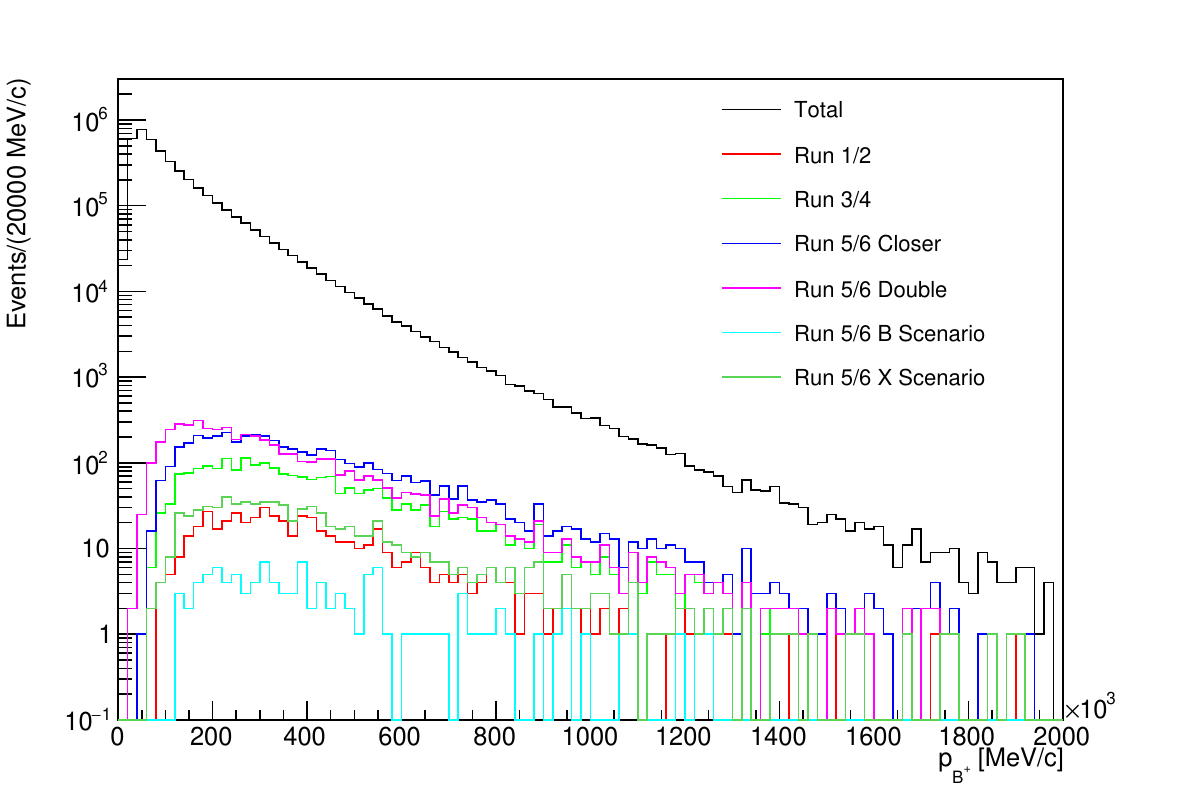}
  \caption{Characteristics of \Bu mesons that produce at least two hits in the detector stations for the different geometry configurations: (top left) flight distance from the primary vertex, (top right) \Bu meson lifetime and (bottom) momentum.}\label{fig:bproperties}
  \end{center}
 \end{figure}

In Figures~\ref{fig:bproperties}, the flight distance, meson lifetime, and meson momentum are shown before and after requiring that the \B meson left two hits in the tracking stations, in the different geometries. 
As expected, most of the reconstructed \Bu mesons have long lifetime and/or high boost due to momentum. 
These are however also characteristics that improve the trigger and selection efficiencies of a \Bu decay analysis, hence 
the total efficiency for analyses using $B$-tracks will be higher than naively expected from multiplying the usual efficiencies with the additional requirement of a reconstructed $B$-track.
This effect will increase the efficiency of this method at high transverse momentum, making it viable for the typical \Bu mesons analysed by multi-purpose experiments such as ATLAS and CMS. 
While the \Bu mesons considered here come from inclusive $pp\to b\bar b$ production, highly boosted \Bu from electroweak or beyond the Standard Model high mass particles could benefit even more from \Bu tracking.
Future higher energy experiments, could also have higher efficiencies due to an even higher boost, 
provided they could keep detectors at the same distance from the primary interaction, and possibly extend the method to $D^+_{(s)}$ mesons. 

Considering specific $\Bu$ decays, while a penalty of $10^{-3}-10^{-4}$ in efficiency could seem large, single event sensitivities of $10^{-11}-10^{-12}$ are already attainable with current datasets at LHC. Therefore, branching fractions of the order of $10^{-7}-10^{-9}$ could be reachable with this method, 
opening up the possibility of reconstructing previously unreconstructible decays at hadron-colliders in a healthy competition with $B$-factory experiments.

\cleardoublepage
 \appendix
 \section{Kinematics of the \Bu decays}
 \label{app}

\begin{figure}[!h]
\begin{center}
 \includegraphics[width = 0.45\textwidth]{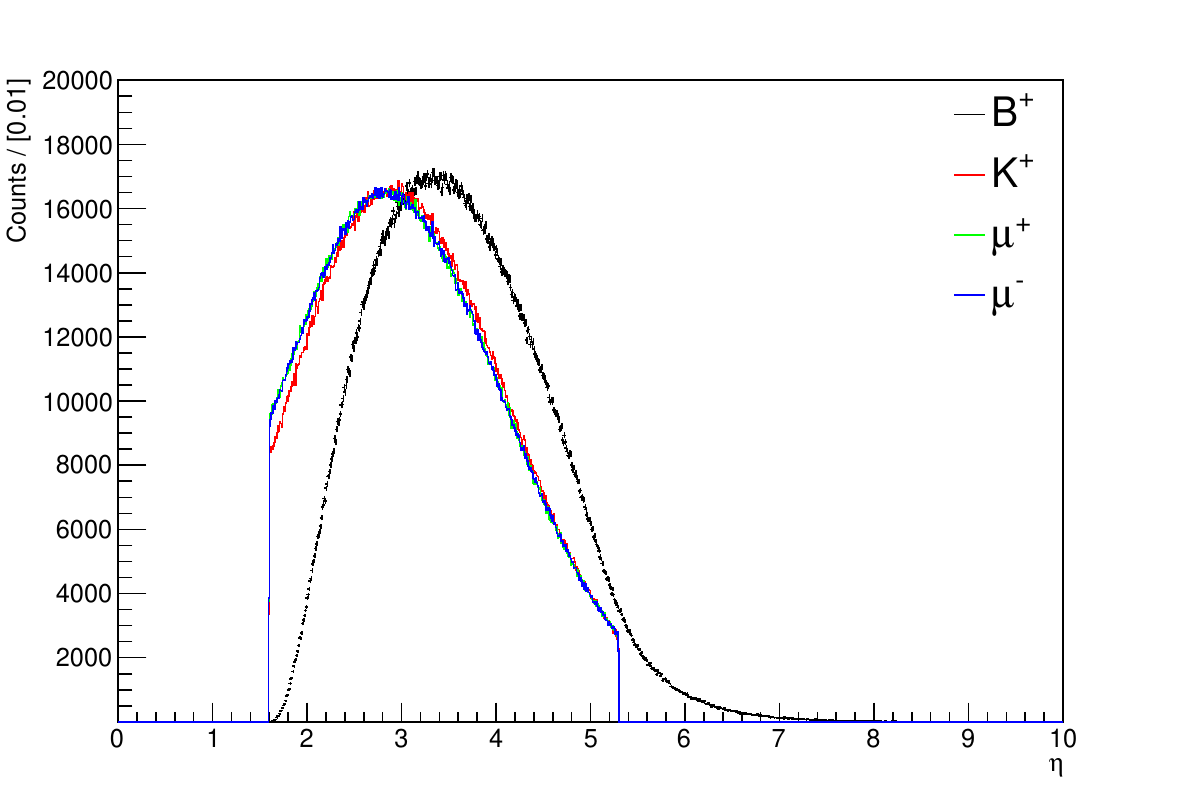}
 \includegraphics[width = 0.45\textwidth]{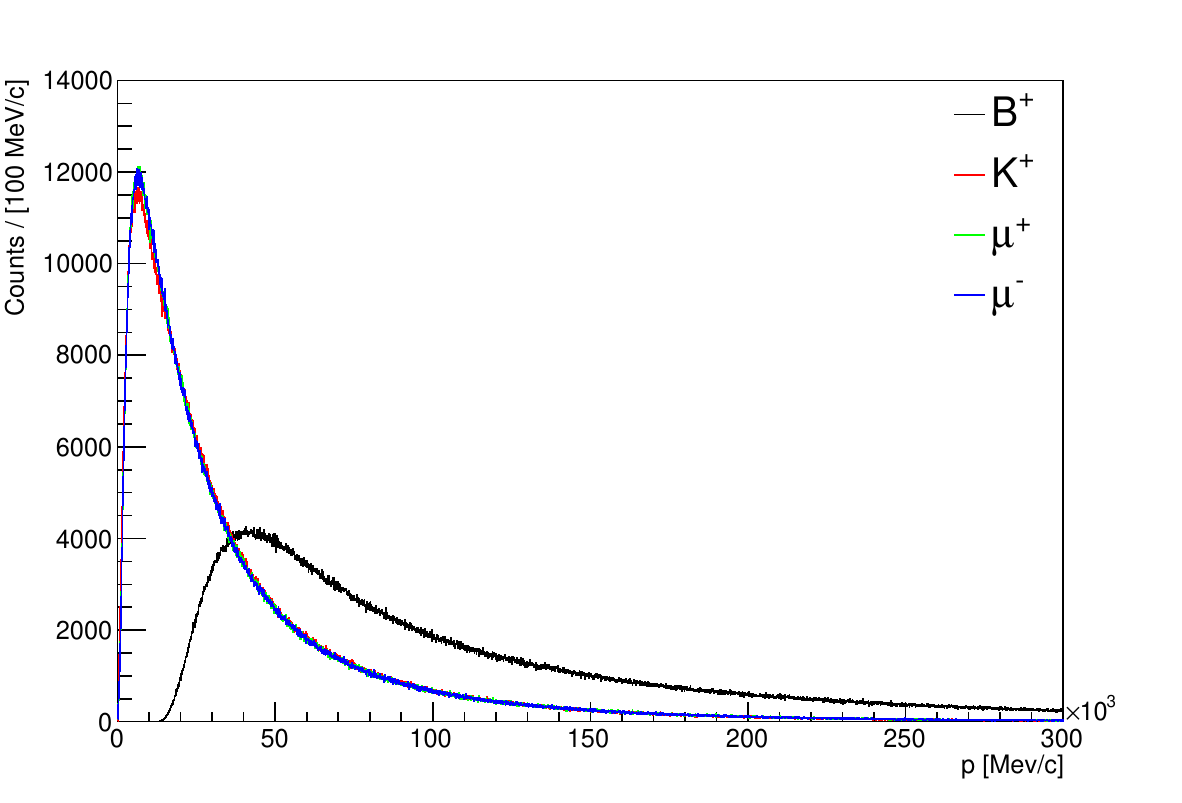}
 \includegraphics[width = 0.45\textwidth]{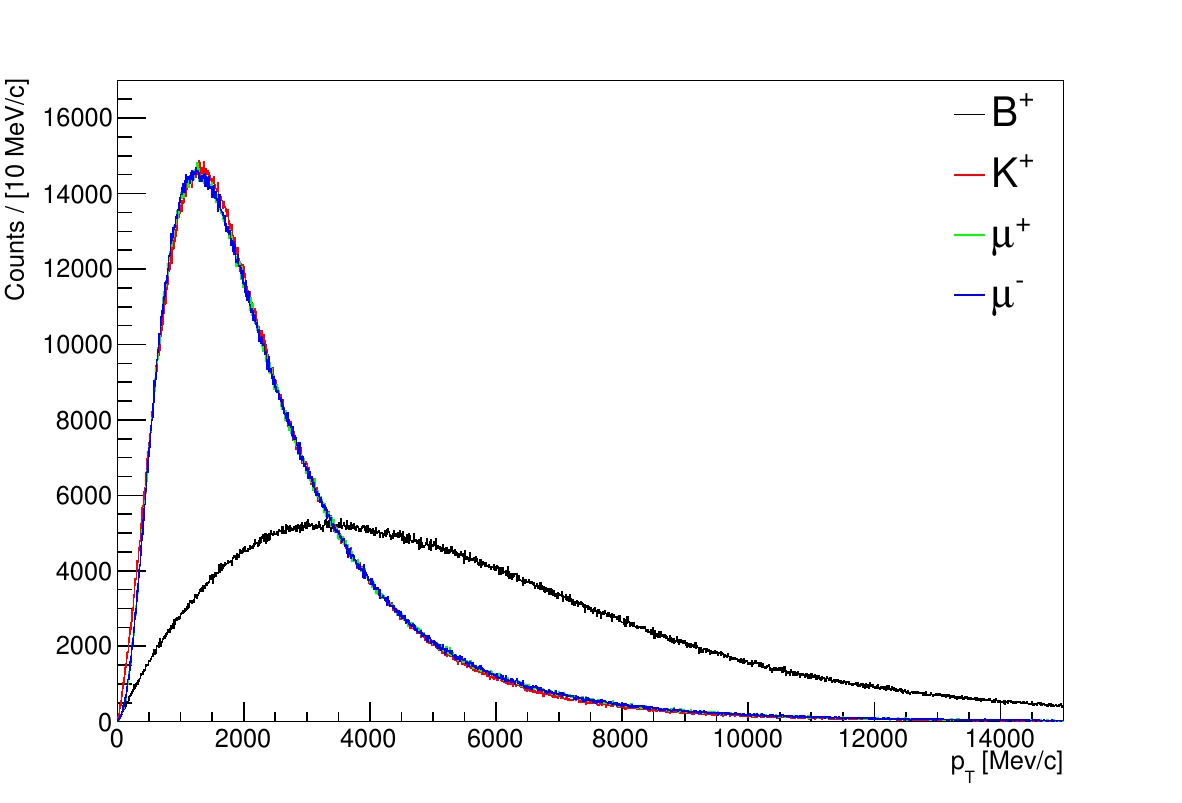} 
  \caption{\emph{Kinematic variables of all the generated particles: pseudorapidity (top left), momentum (top right), and transverse momentum (bottom).}}\label{fig:kin}
  \end{center}
 \end{figure}